\documentclass [10pt,twoside]{article}

\usepackage{shrthnds}
\usepackage{cite}
\usepackage{graphicx}
\usepackage[usenames]{color}
\usepackage{subfigure}
\usepackage{setspace}

\usepackage{multirow}

\usepackage[hang,small]{caption}

\usepackage{amsmath}                  

\usepackage{geometry}
\usepackage{fancyhdr}
\usepackage{titlesec,titletoc}
  \titleformat{\section}{\Large\sf\bfseries}{\thesection}{1em}{}
  \titleformat{\subsection}{\large\sf\bfseries}{\thesubsection}{1em}{}

\title{\sf\bfseries \ntitle}
\author{  Pankaj Jain$^{1}$\footnote{email: pkjain@iitk.ac.in}~,
Purnendu Karmakar$^1$\footnote{email: purnendu@iitk.ac.in}~,
 Subhadip Mitra$^{2}$\footnote{email: smitra@iopb.res.in}~,\\
 Sukanta Panda$^3$\footnote{email: sukanta@iiserbhopal.ac.in}
and Naveen K. Singh$^4$\footnote{email: naveenks@prl.res.in}}
\date{}

\newcommand{\pghdr}{\footnotesize {P. Jain} {\it et al.} -- Testing Unimodular Gravity}
\newcommand{\ntitle}{Testing Unimodular Gravity }

\begin{document}
\vspace{-3cm}
\maketitle
\vspace{-0.6cm}
\bc
\small{ 1) Department of Physics, IIT Kanpur, Kanpur 208 016, India\\
2) Institute of Physics, Bhubaneswar 751 005, India\\
3) Indian Institute of Science Education and Research, Bhopal 462 023, India\\
4) Physical Research Laboratory, Ahmedabad 380009, India}
\ec
\vspace{0.6cm}

\bc\begin{minipage}{0.9\textwidth}\begin{spacing}{1}{\small {\bf Abstract:}
We consider models of gravitation that are based on unimodular general 
coordinate transformations (GCT). These transformations include only those 
which do not change the determinant of the metric. We treat the determinant
as a separate field which transforms as a scalar under unimodular GCT. 
We consider a class of such theories. In general, these theories do not
transform covariantly under the full GCT. We characterize the violation of 
general coordinate invariance by introducing a new parameter. We show that 
the theory is consistent with observations for a wide range of this 
parameter. This parameter may serve as a test for possible violations of 
general coordinate invariance. We also consider the cosmic evolution within
the framework of these models. We show that in general we do not obtain
consistent cosmological solutions if we assume the standard cosmological
constant or the standard form of non-relativistic matter. We propose a 
suitable generalization which is consistent with cosmology. We fit
the resulting model to the high redshift supernova data. We find that 
we can obtain a good fit to this data even if include only a single 
component, either cosmological constant or non-relativistic matter.
}\end{spacing}\end{minipage}\ec


\section{Introduction}
The idea of unimodular gravity was first proposed by Anderson and Finkelstein
\cite{Anderson}. It is closely related to an earlier proposal by Einstein
\cite{Einstein}. In Ref. \cite{Anderson} the authors argued that it is natural to split the metric into two 
separate pieces, its determinant $g$ and the remaining metric whose 
determinant is fixed to unity. In their proposal the determinant of
the metric is not a
field and hence not a dynamical variable. Later, it has been suggested 
that since the determinant is not a field in this theory, the cosmological
constant term does not appear in the action. Thus this might help in 
solving the fine tuning problem of the cosmological constant. However,
as nicely explained in the review paper by Weinberg \cite{Weinberg}, 
the problem is 
not really solved. The full action still has general coordinate invariance 
(GCI).
Using this fact, Weinberg \cite{Weinberg} shows that one again recovers the standard
Einstein's equations. The only difference is that the cosmological constant
now  appears as an integration constant rather than a term in the action. 
The trace of the energy momentum tensor still contributes to the
cosmological constant and the fine
tuning problem is present.

Unimodular gravity has been pursued in detail in many papers 
\cite{Ng1,Zee,Buchmuller,Henneaux89,Unruh1989,Ng2,Finkelstein,Alvarez05,Alvarez06,Abassi,Ellis},
which have considered its application to the problem of the 
cosmological constant and its quantization. 
An interesting generalization is to consider the determinant as a dynamical
field, as suggested by Zee \cite{Zee} and Buchmuller and Dragon 
\cite{Buchmuller}. In this case the determinant effectively acts as a 
scalar field since one only demands
invariance under unimodular general coordinate transformations (GCT). Hence 
one may include the determinant as a scalar field in the
action \cite{Buchmuller,Weinberg} 
and allow the full GCI to be broken. Such a procedure has been pursued 
in some papers in the literature \cite{Buchmuller,Alvarez07,JMS,Alvarez09,Alvarez10,Shaposhnikov}. 
In the present paper we study some implications of this proposal. We
study a model which represents a minimal generalization of the Einstein's 
gravity and hence can be utilized to test the GCI. We introduce a parameter, which
characterizes the violation of GCI and whose 
value may be determined or constrained observationally. For this 
purpose we study the cosmological implications of this model as well
as its spherically symmetric solution in vacuum.
We also use this model to fit the high $z$ supernova data. 
The precise model we consider has not been 
investigated so far in the literature.

\section{Models of Unimodular Gravity}

We propose to split the standard metric as follows,
\begin{eqnarray}
g_{\mu\nu}&=&\chi^2 \bar{g}_{\mu\nu},
\label{eq:gmunu}
\end{eqnarray}
where we impose the following constraint on the determinant of $\bar{g}_{\mu\nu}$,
\begin{eqnarray}
\bar g = det[\bar {g}_{\mu\nu}]= f(x) .
\label{eq:constraint}
\end{eqnarray} 
Here $f(x)$ is some specified function of the space-time coordinates. In
literature this has generally been taken to be unity. However if we fix it
as such then it excludes even the Lorentz metric in spherical coordinates.
It seems better to generalize this constraint to at least allow $f(x)$ to
be the determinant of the Lorentz metric, in what ever coordinates we choose
to express it. We point out that $\bar g$ is not a dynamical variable in this
theory. Eq. \ref{eq:gmunu} implies,
\begin{eqnarray}
g^{\mu\nu}&=&\frac{1}{\chi^2} \,\bar{g}^{\mu\nu}.
\end{eqnarray}

We shall demand that our theory is invariant only under unimodular GCT. Under these transformations,
\be
x^\mu \rightarrow x'^{\mu}
\ee  
such that the Jacobian is unity, {\it i.e.}, 
\be
\det \left(\frac{\partial x'^\mu}{\partial x^\nu}\right)=1  .
\ee
The determinant of $g_{\mu\nu}$ and hence the field $\chi$ behaves 
as a scalar under these transformations. 
We treat $\chi$ as an independent scalar field. The basic quantities
such as the connection, curvature tensor etc. split naturally into a
function of $\bar g_{\mu\nu}$ and $\chi$. We find,
\ba
\Gamma^\mu_{\alpha\beta} = \bar \Gamma^\mu_{\alpha\beta} +\tilde
\Gamma^\mu_{\alpha\beta}\,,
\label{eq:Gamma}
\ea
where $\bar \Gamma^\mu_{\alpha\beta}$ is the connection computed using the 
metric $\bar g_{\mu\nu}$ and 
\be
\tilde\Gamma^\mu_{\alpha\beta} = \bar g^\mu_\beta\, \partial_\alpha \ln\chi
 + \bar g^\mu_\alpha\, \partial_\beta \ln\chi - \bar g_{\alpha\beta}\,
\partial^\mu \ln\chi .
\label{eq:tilde_Gamma}
\ee
We point out that all raising and lowering of indices is done using the 
metric $\bar g_{\mu\nu}$. Furthermore all covariant derivatives are defined
with respect to this metric. We see from Eq. \ref{eq:tilde_Gamma} that
$\tilde\Gamma^\mu_{\alpha\beta}$ behaves as a tensor under unimodular
GCT.
The Ricci curvature tensor may also be written as,
\ba
R_{\mu\nu} = \bar R_{\mu\nu} + \tilde R_{\mu\nu}\,,
\ea
where $\bar R_{\mu\nu}$ is the Ricci tensor computed by using the metric
$\bar g_{\mu\nu}$ and 
\be
\tilde R_{\mu\nu} = -\tilde \Gamma^\alpha_{\mu\nu;\alpha}
+ \tilde \Gamma^\alpha_{\mu\alpha;\nu} - \tilde \Gamma^\alpha_{\beta\alpha}
\tilde \Gamma^\beta_{\mu\nu} + \tilde \Gamma^\alpha_{\beta\nu}
\tilde \Gamma^\beta_{\mu\alpha}.
\ee
We find 
\be
\tilde R_{\mu\nu} = 2(\ln\chi)_{;\mu;\nu} + \bar g_{\mu\nu}
(\ln\chi)^{;\alpha}_{;\alpha} - 2\partial_\mu\ln\chi \,
\partial_\nu\ln\chi + 2 \bar g_{\mu\nu} \partial_\beta\ln\chi\, 
\partial^\beta\ln\chi .
\ee
Contracting both $\bar R_{\mu\nu}$ and $\tilde R_{\mu\nu}$ with the
metric $\bar g^{\mu\nu}$ we obtain the scalars $\bar R$ and $\tilde R$ 
respectively. Hence we can express the Ricci scalar $R$ as
\be
R = g^{\mu\nu} R_{\mu\nu} = \frac1{\chi^2}(\bar R + \tilde R).
\ee 
This gives,
\be
\tilde R = 6 (\ln\chi)^{;\mu}_{;\mu} + 6\partial_\mu \ln\chi
\ \partial^\mu\ln\chi .
\ee 

We may now express the Einstein action in terms of quantities computed
by the metric $\bar g_{\mu\nu}$ and $\chi$,
\be
S_{\rm E} = \int d^4x \sqrt{-\bar g} {1\over 16 \pi G}
\, \left[\chi^2 \bar R + \chi^2 \tilde R\right].
\ee
Expressing $\tilde R$ explicitly in terms of $\chi$ we find, 
after an integration by parts,
\be
S_{\rm E} = \int d^4x \sqrt{-\bar g} \frac{1}{16 \pi G}
\, \left[\chi^2 \bar R - \xi\partial_\alpha\chi \partial^\alpha\chi\right],
\label{eq:SE}
\ee
where the parameter $\xi=6$. 
This action, given in Eq. \ref{eq:SE}, is exactly the same as the 
standard Einstein's action as long as $\xi = 6$. 
However we treat $\xi$ as an arbitrary parameter to be fitted to experimental
data. For other values of $\xi$, the action in  Eq. \ref{eq:SE} will respect
only unimodular GCI but not the full symmetry group.  
Hence the parameter $\xi$ also provides a test of the validity of the GCI.

The action given in Eq. \ref{eq:SE} has to be supplemented by the matter action. We discuss
some simple examples of matter action in the next section. Here we clarify that our model
preserves most of the structure of Einstein's theory. Hence we still have the freedom to 
choose a locally inertial coordinate system which is valid in a small neighborhood of
any point. This is because our metric $\bar g_{\mu\nu}$ has determinant equal to the 
determinant of the Lorentz metric. Hence we can transform it locally to the Lorentz
metric by making a unimodular GCT. Indeed we may choose any coordinate system
in our unimodular theory subject to the constrain that we can locally 
transform it to the Lorentz metric by making a unimodular GCT. 

There are many possible ways to generalize the model such that it displays
invariance under unimodular GCT but not the full GCI. We may, for example,
include terms with any power of the field $\chi$ without breaking 
unimodular GCI 
\cite{Buchmuller,Alvarez07,JMS,Alvarez09,Alvarez10,Shaposhnikov}. 
We may also introduce other terms which violate GCI \cite{Anber10}.
The model presented in Eq. \ref{eq:SE} with $\xi\ne 6$
represents a minimal modification of the Einstein's gravity. 
The matter
action may also be modified so as to display only unimodular GCI. 
We shall consider some simple examples of such modification in the next 
section. 
It may be interesting to consider the most 
general model within the framework of unimodular gravity. However such a
model would introduce a large number of parameters and it appears better
to first get some familiarity with these models in a simpler framework.  

We point out that our theory is covariant under unimodular GCT. 
Hence the action and the field equations
remain invariant under these transformations. Under  
the full GCT, the action and the equations
of motion would change. However one may make the action look formally
invariant under the full GCT by introducing
 a compensator field \cite{Alvarez07}. Let us assume that we define our
unimodular theory in some coordinate system $\bar x$. One may, of course,
choose any coordinate system which is related to $\bar x$ by a
unimodular transformation. Now lets make a general coordinate transformation,
denoting the new coordinates by $x$. The compensator field is defined as, 
$C(x) \equiv D(x,\bar x)$, 
which is the determinant of the transformation between  
$\bar x$ and $x$ \cite{Alvarez07}. 
In this case the action formally appears invariant under GCT \cite{Alvarez07,Shaposhnikov}, 
with the introduction
of an additional scalar field. This procedure may offer some advantages 
for some applications. However in the present case we prefer to work
directly with the original variables without introducing the compensator
field. Let us consider, for example, an
application of our model to cosmology. Here the simplest metric to use is 
precisely a preferred frame in which we expect our dynamical equations
to reduce to the unimodular equations. We discuss this in section 3. 
Hence it appears simpler to work
directly with the original variables. Furthermore we point out that
one can use the general coordinate system, with suitably modified action,
only to compute observables which are invariant under GCT. 
However in a unimodular theory all quantities which
are invariant under a limited unimodular transformations are physically
observable. For quantities which are invariant under only the unimodular
transformations and not the full GCT, the results will depend on the 
coordinate system, if we use the covariant construction by introducing a
compensator field. Hence in this sense the unimodular theory in not completely
equivalent to the fully covariant theory constructed by introducing a 
compensator field. For these reasons,
 here we do not introduce the compensator field
and work directly with the original non-covariant theory. 

\subsection{Particle Trajectory Equation}
The geodesic equation for particle motion can be expressed in terms of
the variable $\chi$ by use of Eq. \ref{eq:Gamma}. We find,
\be
{d^2 x^\mu\over d\lambda^2} + \bar\Gamma^\mu_{\alpha\beta} {dx^\alpha\over
d\lambda} {dx^\beta\over d\lambda}
+ \tilde\Gamma^\mu_{\alpha\beta} {dx^\alpha\over
d\lambda} {dx^\beta\over d\lambda} = 0.
\label{eq:geodesic}
\ee 
The first two terms on the left hand side represent the usual terms due
to motion of particle in gravitational field described by the metric
$\bar g_{\mu\nu}$. The third term represents the additional contribution due
to $\chi$.  Hence
the scalar field $\chi$ simply provides an additional force which the particles
experience. If we demand only unimodular GCI it is of course possible to
generalize this equation.  
We may, for example, insert a parameter multiplying
the last term on the left hand side of this equation. Limits on this
parameter may be imposed observationally. In this paper we shall assume
that this parameter is unity. In the next section we discuss cosmic evolution within
the framework of our model. As we shall see this requires modification of the
particle trajectory equation in some cases.

\section{Cosmic Evolution}
We next use our model to consider evolution of the universe.
The unimodular gravitational equation of motion may be written as
\ba
-\chi^2 \left[\bar R_{\mu\nu} - {1\over 4} \bar g_{\mu\nu}\bar R\right]
- \left[\left(\chi^2\right)_{;\mu;\nu} - {1\over 4}\bar g_{\mu\nu} 
\left(\chi^2\right)^{;\lambda}_{;\lambda} \right]
+\xi\left[\partial_\mu\chi\partial_\nu\chi -{1\over 4}\bar g_{\mu\nu}
\partial^\lambda\chi\partial_\lambda\chi \right]\nonumber\\
 = {\kappa\over 2} \left[T_{\mu\nu}
- {1\over 4} \bar g_{\mu\nu} T^\lambda_\lambda\right] ,
\label{eq:Rmunu}
\ea
where $\kappa = 16\pi G$. 
The equation of motion of the field $\chi$ gives,
\be
2 \chi\bar R + 2\xi \bar g^{\mu\nu} \chi_{;\mu;\nu} = \kappa T_\chi,
\label{eq:chi}
\ee
where $T_\chi$ includes all the contribution due to the coupling of
$\chi$ with matter fields. We shall discuss examples of this below.
Here we may point out that the energy-momentum tensor, $T_{\mu\nu}$, as defined in
Eq. \ref{eq:Rmunu}, 
need not satisfy the usual conservation law. It is simply the contribution
given by matter fields to this equation. The curvature tensor does indeed
satisfy the Bianchi identity but this does not necessarily imply a 
conservation law for $T_{\mu\nu}$. We may, however, obtain a conservation
law by obtaining an expression for 
$\bar R_{\mu\nu} -  \bar g_{\mu\nu}\bar R/2$
by using Eqs. \ref{eq:Rmunu} and \ref{eq:chi}.
This will define a generalized
energy momentum tensor which will satisfy the conservation law,
since,
\be
\left[\bar R^\mu_\nu - {1\over 2} \bar g^\mu_\nu\bar R\right]_{;\mu} = 0.
\label{eq:conservation}
\ee

\subsection{Cosmological Constant}
We first consider the simple case of cosmological constant or
 vacuum energy. This term contributes 
to the action in the form
\be
S_{vac} = \int d^4x \sqrt{-\bar g}\left[ - \chi^4\Lambda\right],
\label{eq:Svac}
\ee
which gives,
\be
T_\chi = 4\chi^3\Lambda .
\label{eq:Tchi_vac}
\ee
However it does not contribute to Eq. \ref{eq:Rmunu}. For simplicity we 
assume a spatially flat FRW metric,
\be
g_{\mu\nu} = \chi^2(\eta)\bar g_{\mu\nu}
\label{eq:metricFRW}
\ee
with
\be
\bar g_{\mu\nu} = {\rm diagonal}\,[1,-1,-1,-1].
\ee
Here $\eta$ denotes the conformal time. Our metric is essentially
the same as the standard spatially flat Friedmann-Robertson-Walker (FRW) metric written in terms
of conformal time.  
We do not consider more complicated metrics in the present paper. 
This implies $\bar R=0$ and $\bar R_{\mu\nu} = 0$. 
Setting $\mu=\nu=0$ in Eq. \ref{eq:Rmunu} we find
\be
2\chi {d^2\chi\over d\eta^2} + (2-\xi)\left[ {d\chi\over d\eta}\right]^2 = 0.
\label{eq:vacuum1}
\ee
Using Eqs. \ref{eq:chi} and \ref{eq:Tchi_vac} we find 
\be
2\xi {d^2\chi\over d\eta^2} = 4\kappa\chi^3\Lambda
\ee
This equation along with Eq. \ref{eq:vacuum1} leads to, 
\be
{d\chi\over d\eta} = \pm \sqrt{4\kappa\Lambda\over \xi(\xi-2)}\,\chi^2.
\label{eq:vacuum2}
\ee
For $\xi = 6$ this gives the standard solution obtained
by using the Einstein's equations. For $\xi\ne 6$ we find that 
there does not exist any solution to these set of equations, 
Eq. \ref{eq:vacuum1} and Eq. \ref{eq:vacuum2}. The 
only allowed solution is that $\chi=0$. Hence we find
a very interesting result that the vacuum dominated solution  
exists only for the standard case $\xi = 6$. This is interesting 
since it might provide us with a solution to the problem of
fine tuning of the cosmological constant. A large cosmological constant
here does not imply a rapidly expanding universe. A consistent 
solution may be obtained with a cosmological constant only
if some other component, such as matter or radiation, also contributes
to energy density. Alternatively we may generalize the action for
the cosmological constant term, which is permissible within the framework
our unimodular theory. We discuss this in the next subsection.

\subsection{Generalized Cosmological Constant}
We next determine whether it is possible to generalize the cosmological constant
term such that it leads to consistent cosmology in the general case when
$\xi$ is different from 6. We assume an action of
the form
\be
S'_{vac} = \int d^4x \sqrt{-\bar g}\left[ - \chi^\delta\Lambda\right],
\label{eq:Spvac}
\ee
instead of Eq. \ref{eq:Svac}. In Eq. \ref{eq:Spvac} 
$\delta$ is a constant. We shall fix this constant by demanding that 
it leads to consistent cosmology. For the standard case of $\xi=6$ we
have $\delta=4$.  
The unimodular gravitational equation of motion remains the same as
Eq. \ref{eq:vacuum1} whereas the equation of motion for $\chi$ changes to, 
\be
2\xi {d^2\chi\over d\eta^2} = \delta\kappa\chi^{(\delta-1)}\Lambda
\label{eq:vacuum3}
\ee
Eliminating the second derivative between Eq. \ref{eq:vacuum1} and Eq. \ref{eq:vacuum3},
we obtain
\be
{d\chi\over d\eta} = \pm \sqrt{\delta\kappa\Lambda\over \xi(\xi-2)}\,\chi^{\delta/2}.
\label{eq:vacuum4}
\ee
Demanding that this equation is consistent with Eq. \ref{eq:vacuum3} we obtain,
\be
\delta = \xi -2
\label{eq:cond1}
\ee
We next solve for $\chi(t)$ where $t$ is the cosmic time such that $\chi d\eta = dt$. 
The expanding solution, corresponding to the positive sign in 
Eq. \ref{eq:vacuum4}, 
is found to be
\be
\chi(t) = \left[1-{6-\xi\over 2} \sqrt{\kappa\Lambda\over \xi} (t_0-t)\right]^{2/(6-\xi)}
\label{eq:solnvac}
\ee
Here $t_0$ is the current time and we have set $\chi(t_0) =1$. 

It is useful to determine whether this vacuum component by itself can 
explain cosmological observations. We partially address this issue in 
the present paper by fitting this model to the high redshift supernova data.
We next obtain the luminosity distance, $d_L$, in this model.
  The luminosity distance is given by, 
\begin{equation}
d_L=\frac{r}{\chi}
=(z+1)\int\limits_{t(z)}^{t_0}\frac{dt'}{\chi(t')}\,,
\end{equation}
where $t_0$ is the current time and $t(z)$ is the time when the light
left the source located at a redshift of $z$.
We obtain
\be
d_L = {2(1+z)\over (\xi-4) H_0}\left[(1+z)^{(\xi-4)/2} - 1\right]
\label{eq:dL_vac}
\ee
where $H_0= \sqrt{\kappa\Lambda/\xi}$ is the current value of the Hubble
constant.
The relation between luminosity distance and distance modulus $\mu$ is
given by, 
\begin{eqnarray}
\mu= m-M &=& 5(\log_{10} d_L-1)\,.
\end{eqnarray}

\begin{figure}[!ht]
\begin{center}
\includegraphics[width=0.75\textwidth]{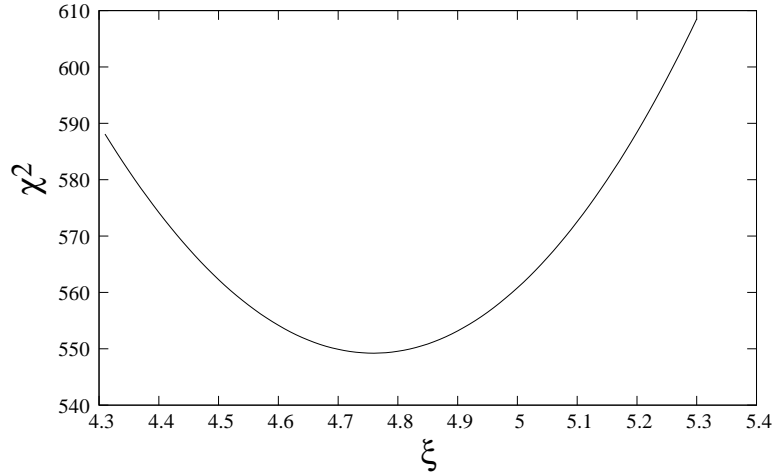}
\end{center}
\caption{The $\chi^2$ as a function of the parameter $\xi$ for purely vacuum
dominated Universe as defined by Eq. \ref{eq:Spvac}.}
\label{fig:chi2_vac}
\end{figure}

We fit the supernova data using the compilation of 557 sources
available in \cite{Amanullah}. The resulting statistic $\chi^2$ as a
function of the parameter $\xi$ is shown in Fig. \ref{fig:chi2_vac}. 
Here we have 
used the $d_L$ given by Eq. \ref{eq:dL_vac} obtained by including only
the vacuum contributions defined by the action, Eq. \ref{eq:Spvac}. 
We find that $\chi^2$ attains its minimum value of 549.2 at $\xi=4.76$. 
At this value of $\xi$ the Hubble parameter is found to be
69.4 Km/(sec Mpc). Since $\chi^2$ per degree of freedom is less than
unity we find that this single component model also provides a good
fit to the data. For comparison the standard $\Lambda$CDM
model, which includes both cosmological constant and non-relativistic
dark matter, leads to $\chi^2$ equal to 542.8. 

\subsection{Radiation}
We next solve our equations for a universe dominated by relativistic matter.
We assume a scalar field with action given by,
\be
S_{rad} = \int d^4x \sqrt{-\bar g} {1\over 2}\chi^2\bar g^{\mu\nu} \partial_\mu\phi
\partial_\nu\phi,
\ee
where the mass term for the scalar $\ph$ is assumed to be negligible. We identify energy momentum
tensor as,
\be
T_{\mu\nu} = \chi^2 \left<\partial_\mu\phi \partial_\nu\phi\right>,
\label{eq:T_rad}
\ee 
where the expectation value is taken in an appropriate thermal state
corresponding to the temperature of the medium. Here we are treating $\phi$
as a quantum field and the fields $\bar g_{\mu\nu}$, $\chi$
etc. classically. In Eq. \ref{eq:T_rad} the contribution to $T_{\mu\nu}$
is obtained by identifying the contribution from the matter fields to the
unimodular gravitational equations of motion, Eq. \ref{eq:Rmunu}. We do
not get any contributions proportional to $\bar g_{\mu\nu}$, which are
present in the standard energy momentum tensor. In any case, such contributions
would cancel when we subtract the trace.  

We can now relate $\left<\partial_\mu\phi \partial_\nu\phi\right>$ to the usual definition
of energy density. Let $\Theta_{\mu\nu}$ denote the standard energy 
momentum tensor in the Einstein's gravity. Then the standard expression for
energy momentum tensor gives,
\be
\Theta_{00} = \left<\partial_0\phi\partial_0\phi\right> = g_{00} \Theta^0_0 = a^2(\eta)\rho,
\ee 
where $a(\eta)$ is the standard FRW scale factor defined in terms of conformal
time and $\rho$ is the energy density of radiation. Here we have used the
fact that $\Theta_0^0=\rho$. Furthermore, we identify the scale factor 
$a(\eta)$ with $\chi$. Hence we obtain,
\be
\left<\partial_0\phi\partial_0\phi\right> = \chi^2(\eta)\rho.
\label{eq:radrho}
\ee 
The contribution $T_\chi$ to Eq. \ref{eq:chi} is given by 
\be
T_\chi = -\chi \left<\bar g^{\mu\nu} \partial_\mu\phi\partial_\nu\phi\right>.
\ee
In momentum space each particle will contribute a term proportional to 
$\bar g^{\mu\nu}P_\mu P_\nu$,
where $P_\mu$ is its momentum, to this equation. 
Hence $T_\chi$ is equal to zero for relativistic matter.  
 
We next solve the equations of motion setting $\bar g_{\mu\nu}$ equal to the
Minkowski metric. 
The equation of motion for $\chi$, Eq. \ref{eq:chi}, gives
\be
{d^2\chi\over d\eta^2}  = 0\ .
\ee
 This equation implies
that $d\chi/d\eta={\rm constant}$. The unimodular
equations of motion, Eq. \ref{eq:Rmunu} gives
\be
(\xi-2) \left({d\chi\over d\eta}\right)^2 = {2\kappa\over 3} \chi^4\rho.
\ee
This gives the standard equation for the scale parameter in terms of
conformal time if we set $\xi=6$. For other values it represents
the generalization to unimodular gravity. Irrespective of the value
of $\xi$ we find that
\be
\rho \propto {1\over \chi^4}.
\label{eq:rho_rad}
\ee
The value of $\xi>2$ since for smaller values the equation gives complex
solutions, which are physically not acceptable. For all values of 
$\xi>2$, the evolution is the same as in the standard Einstein's gravity
up to an overall constant factor which may be absorbed in the initial
condition on the energy density of the radiation field.

We may verify the relationship between $\rho$ and $\chi$, Eq. \ref{eq:rho_rad}, 
directly from the action. For this purpose we define the scaled variable 
$\bar \phi = \chi\phi$. In terms of the scaled field we may write the action,
after an integration by parts, as,
\be
S_{rad} = \int d^4x \sqrt{-\bar g} {1\over 2}\bar g^{\mu\nu} 
\partial_\mu\bar\phi\partial_\nu\bar\phi + 
\int d^4x \sqrt{-\bar g} {1\over 2}\bar g^{\mu\nu} 
{\bar\phi^2\over \chi}\partial_\mu\partial_\nu\chi ,
\ee
We assume that the background field $\chi$ is
a slowly varying function of time. Hence in the adiabatic limit we 
may ignore its derivatives. Within this approximation we may write the
action as
\be
S_{rad} \approx \int d^4x \sqrt{-\bar g} {1\over 2}\bar g^{\mu\nu} 
\partial_\mu\bar\phi
\partial_\nu\bar\phi,
\ee
The corrections to this are of order $H^2/\omega^2$, where $H$ is the Hubble
constant and $\omega$ the frequency of the radiation field.
 Hence for 
all frequencies of physical interest, this ratio is extremely small and 
can be safely neglected. We may now treat this as the standard massless
free field in flat space-time since we have taken $\bar g_{\mu\nu}$
as the Lorentz metric.
The expectation value of the Hamiltonian density,
$\cal H$, is simply equal to the energy density $E/V$ where $E$ is the total
energy and $V$ the spatial volume. For 
any multi-particle eigenstate of the Hamiltonian this is equal to the 
sum of frequencies of
all the particles divided by the volume of space. We point out that 
the conformal time, $\eta$, dependence of the free field solutions
is given by $e^{-i\eta\omega}$. The spatial volume in the present
flat space-time case is independent of $\chi$. The frequencies, $\omega$,
are also independent of $\chi$. Hence we find,
\be
\left<\cal H\right> \approx \left<{\partial\bar\phi\over \partial\eta}
{\partial\bar\phi\over \partial\eta}\right> 
\label{eq:exH}
\ee 
is independent of $\chi$. This is true for all states since they can 
be written as linear superpositions of the eigenstates. We point out that the
two terms involving the time and space derivatives in the Hamiltonian
density contribute equally to the expectation value in Eq. \ref{eq:exH}. 
This implies that 
\be
 \left<{\partial\phi\over \partial\eta}
{\partial\phi\over \partial\eta} \right> \propto {1\over \chi^2} 
\ee
Finally using Eq. \ref{eq:radrho} we obtain Eq. \ref{eq:rho_rad}.  

\subsection{Non-relativistic Matter}
For the non-relativistic matter we again use a
scalar field to obtain the matter contributions to our equations of motion.
We first use the standard covariant action decomposed in terms of
the field $\chi$ and the metric $\bar g_{\mu\nu}$. As we shall see this will
not lead to consistent cosmology as we found in the case of cosmological 
constant. We shall then propose a suitable modification by demanding 
consistency with cosmological evolution.
The action may be written as
\be
S = \int d^4x \sqrt{-\bar g} \left[{1\over 2}\chi^2\bar g^{\mu\nu} \partial_\mu\phi
\partial_\nu\phi - {1\over 2} m^2\chi^4\phi^2\right],
\label{eq:Ascalar_mass}
\ee
where, as before, the matter action is same as that in the Einstein's gravity.
We have
\ba
T_\chi &\!=\!& -\chi \left<\bar g^{\mu\nu} \partial_\mu\phi\partial_\nu\phi\right>
+ 2\chi^3 \left<m^2\phi^2\right>,\\
T_{\mu\nu} &=& \chi^2 \left<\partial_\mu \phi\partial_\nu\phi\right> .
\ea
As in the case of radiation we again have,
\be
\left<\partial_0\phi \partial_0\phi\right> = \chi^2\rho,
\label{eq:rho_NR}
\ee
where $\rho$ now stands for the energy density of the non-relativistic matter.
Furthermore we have $\left<m^2\phi^2\right>=\rho$. We again solve the equations with
$\bar g_{\mu\nu}$ equal to the Minkowski metric.
The equation of motion for $\chi$,
Eq. \ref{eq:chi}, gives  
\be
2\xi {d^2\chi\over d\eta^2} = \kappa \chi^3\rho.
\label{eq:NRchi}
\ee
The modified Einstein's equation, Eq.  \ref{eq:Rmunu}, gives,
\be
- {2\over \chi} {d^2\chi\over d\eta^2} + 
(\xi-2){1\over \chi^2} \left({d\chi\over d\eta}\right)^2
= {\kappa\over 2}\chi^2\rho
\label{eq:ModifiedEinNR}
\ee
We eliminate $\rho$ between this equation and Eq. \ref{eq:NRchi} to obtain,
\be
(2+\xi) \chi {d^2\chi\over d\eta^2} = (\xi-2)\left({d\chi\over d\eta}\right)^2.
\ee
This gives,
\be
{d\chi\over d\eta} \propto \chi^\alpha,
\ee
where $\alpha = (\xi-2)/(\xi+2)$. As expected, this gives the standard
evolution for non-relativistic matter if we set $\xi=6$. 
However the evolution changes for $\xi\ne 6$. This is in contrast to the
result we obtained in the case of vacuum dominated or radiation dominated
universe. The evolution is well defined as long as $\xi>2$. 
The evolution of the energy density in this case is given by,
\be
\rho\propto {\chi^{2\alpha}\over \chi^4}
\label{eq:rho_NR1}
\ee
For $\xi=6$, this gives the standard $1/\chi^3$ behavior. However we find
a different behavior for $\xi\ne 6$. 

Our solution above shows that for $\xi \ne 6$, $\rho$ does not decay 
as $1/\chi^3$. However we would have expected this behaviour since our 
matter action is exactly the same as in the case of standard Big Bang model
based on covariant gravity. 
We next explicitly determine the dependence of $\rho$ on $\chi$ in the 
non-relativistic limit directly from our action. As in the case of radiation
field we work in the adiabatic limit and assume that $\chi$ is a slowly
varying function of $\eta$. We now rescale our field such that
$\bar \phi = \chi\phi$. We also define the time varying mass
 $\bar m = \chi m$. In terms of these variables, we may write the action
as, 
\be
S \approx \int d^4x \sqrt{-\bar g} \left[{1\over 2} \bar g^{\mu\nu} 
\partial_\mu\bar\phi
\partial_\nu\bar\phi - {1\over 2} {\bar m}^2(\bar\phi)^2\right],
\ee
Here we have ignored derivatives of $\chi$ since we assume adiabaticity.
The correction terms are proportional to $H^2/m^2$, where $H$ is
the Hubble constant. Hence, for a wide range of values of $m$, these can
be safely neglected.
In terms of $\bar \phi$ and $\bar m$ the action is same as the standard
free scalar field action in flat space-time. Hence we can directly 
use known results for this theory. Here we are interested in the 
extreme non-relativistic limit and hence the time dependence of free field
solutions is given by, $\bar \phi\sim e^{-i\bar m \eta}$. The 
expectation value of the Hamiltonian density $\cal H$ in any 
energy eigenstate is equal to $E/V$ where $E=N\bar m$ is the energy 
corresponding to that state, $N$ is the number of particles 
and $V$ the spatial volume. The 
expectation value of $\cal H$ gets equal contribution from the kinetic 
energy and potential energy terms. Hence we find,
\be
\left<\cal H\right> \approx \left<{\partial\bar\phi\over \partial\eta}
{\partial\bar\phi\over \partial\eta}\right> \propto {\bar m\over V}
\propto \chi 
\label{eq:exH1}
\ee 
This implies that 
\be
 \left<{\partial\phi\over \partial\eta}
{\partial\phi\over \partial\eta} \right> \propto {1\over \chi} 
\label{eq:parphi}
\ee
Finally using Eq. \ref{eq:rho_NR} we find that $\rho\propto 1/\chi^3$. 
Hence we find that the dependence of $\rho$ on $\chi$, given by Eq. 
\ref{eq:rho_NR1}, is inconsistent with the expected behaviour if $\xi\ne 6$. 
This implies that we do not obtain a consistent cosmic evolution for
$\xi \ne 6$, if we 
assume the usual form of non-relativistic matter. 

\subsection{Generalized Non-relativistic Matter}
We next determine whether it is possible to generalize the action
such that it leads to consistent cosmology in the non-relativistic limit.
We propose a modified free scalar field action of the form
\be
S = \int d^4x \sqrt{-\bar g} \left[{1\over 2}\chi^2\bar g^{\mu\nu} \partial_\mu\phi
\partial_\nu\phi - {1\over 2} m^2\chi^{2+2\zeta}\phi^2\right],
\label{eq:Ascalar_mass1}
\ee
where $\zeta$ is a parameter which we will fix by demanding consistency
with cosmic evolution. 
Following the steps leading to Eq. \ref{eq:parphi} 
in the previous section, we find in the present case
\be
 \left<{\partial\phi\over \partial\eta}
{\partial\phi\over \partial\eta} \right> \propto {1\over \chi^{2-\zeta}} 
\ee
Hence we find,
\be
\rho =  {\rho_0\over \chi^{4-\zeta}}
\label{eq:rhochi3}
\ee
where $\rho_0$ is a constant.
We shall fix the value of $\zeta$ by demanding consistency with cosmic
evolution.
We point out that the particle trajectory equation, Eq. \ref{eq:geodesic},
will get modified in the present case for massive particles. The 
third term in Eq. \ref{eq:geodesic} which involves derivatives of $\chi$ will
change since the mass term in the matter action is different from the
standard covariant action. However it will remain unchanged for massless
particles. Hence the calculation of the luminosity distance, discussed 
below, remains the same as in the case of standard covariant theory.  

The equation of motion for $\chi$ gets modified to
\be
2\xi {d^2\chi\over d\eta^2} = \kappa\zeta \chi^3\rho.
\label{eq:NRchi1}
\ee
whereas the modified Einstein equation remains the same as Eq. 
\ref{eq:ModifiedEinNR}. We find a consistent solution to these two equations
only if
\be
\zeta = {\xi-4\over 2}
\ee
The resulting solution in terms of cosmic time $t$ ($\chi d\eta=dt$)
is found to be,
\be
\chi = \left[1-\beta C_1 (t_0-t)\right]^{1/\beta}
\ee
where $\beta = (12-\xi)/4$ and
\be
C_1^2 = {\kappa\rho_0\over \xi}\ ,
\ee
Here we have set $t_0$ as the current time such that 
$\chi(t_0)=1$. 
The luminosity distance in the present model is found to be
\be
d_L = {(1+z)\over (\beta-1)C_1}\left[1-{1\over (1+z)^{(\beta-1)}}\right]
\ee
where $z$ is the redshift.
A fit to the high $z$ supernova data including only such non-relativistic
matter leads to $\xi=9.52$ with $\chi^2 = 549.2$ The Hubble constant in this
case is found to be 69.4 Km/(sec Mpc). 
Hence in this case also we find
a good fit to data purely with non-relativistic matter, since
$\chi^2$ per degree of freedom less than unity. 
The energy density $\rho$ in this case falls as $1/\chi^{2.69}$.

\section{Schwarzschild Solution}
An important check on the parameter $\xi$ is to test whether it leads to a 
modification of the standard spherically symmetric Schwarzschild solution in vacuum. 
We address this question in this section. Imposing the unimodular constraint
on the metric $\bar g_{\mu\nu}$ we can write it as
\be
\bar g_{\mu\nu} = {\rm diag} \left[{1\over A(r)}, -A(r),-r^2,-r^2\sin^2\theta
\right].
\ee 
The full metric is given in Eq. \ref{eq:gmunu} with $\chi=\chi(r)$. 
The determinant $\bar g$ is equal to the determinant of the Lorentz metric. The curvature tensor
satisfies the following equation in vacuum, 
\be
-\chi^2 \left[\bar R_{\mu\nu} - {1\over 4} \bar g_{\mu\nu}\bar R\right]
- \left[\left(\chi^2\right)_{;\mu;\nu} - {1\over 4}\bar g_{\mu\nu} 
\left(\chi^2\right)^{;\lambda}_{;\lambda} \right]
+\xi\left[\partial_\mu\chi\partial_\nu\chi -{1\over 4}\bar g_{\mu\nu}
\partial^\lambda\chi\partial_\lambda\chi \right] = 0.
\label{eq:Rmunu1}
\ee
The equation of motion for $\chi$ may be written as
\be
\chi\bar R = -\xi \bar g^{\mu\nu}\chi_{;\mu;\nu}\,.
\label{eq:chi1}
\ee 
We have 
\be
{\bar R_{rr}\over A} + A \bar R_{tt} = 0.
\ee
Hence we get 
\be
{1\over A} \left(\chi^2\right)_{;r;r} - {\xi\over A} \partial_r\chi
\partial_r\chi + A \left(\chi^2\right)_{;t;t} = 0.
\ee
This leads to the equation 
\be
\chi {d^2\chi\over dr^2} - {\xi - 2\over 2} \left({d\chi\over dr}\right)^2 = 0.
\ee
This implies that
\be
{d\chi\over dr} = {C \chi^\alpha},
\ee
where $C$ is a constant and $\alpha = (\xi-2)/2$. We impose the boundary
condition that
\be
{d\chi\over dr} \rightarrow 0
\ee
as $r\rightarrow \infty$. We also require that $\chi\ne 0$ as $r\rightarrow\infty$. 
This boundary condition implies that the constant $C=0$. Hence $d\chi/dr=0$
and $\chi$ is equal to a constant which we may set equal to unity. 
From Eq. \ref{eq:chi1} we find that $\bar R=0$. Hence Eq. \ref{eq:Rmunu1}
gives $\bar R_{\mu\nu}=0$. It is clear that this will lead to the standard
equation for $A$ and give the standard Schwarzschild solution.

The equation for particle trajectory, Eq. \ref{eq:geodesic}, 
is also the same as in the case of the Einstein's gravity both
for massless and massive particles. This is because $\chi$ is constant
in the present case.  
Hence the predictions of the standard Schwarzschild solution for
spherically symmetric systems remains the same in our theory.
In particular we shall obtain the standard Newtonian potential.
Furthermore the standard
tests of Einstein's theory, such as the perihelion shift of Mercury, would
be preserved in this theory.
We point out that
in the present case the third term on the left hand side of Eq. 
\ref{eq:geodesic} will have no influence since $\chi$ is constant. Hence
even if we allow an arbitrary parameter multiplying this term, it will
not change the predictions of the standard Schwarzschild solution.

\section{Discussion and Conclusions}
We have discussed a class of models which respect only unimodular GCI but
not the full GCI. In general such models may be constructed by introducing
many new parameters. In the present paper we restrict ourselves to a 
minimal extension by introducing just one parameter, $\xi$, in the gravitational
sector. For 
$\xi=6$, the model reduces to Einstein's gravity. We explore
a range of values of this parameter such that the model leads to
acceptable cosmic evolution. We find that for a wide range of this parameter,
$\xi>2$, the cosmic evolution remains unchanged in the radiation 
dominated epoch. However we do not find a consistent solution in 
the case of vacuum dominated Universe 
for $\xi\ne 6$. Hence if $\xi$ is different
from 6, a large cosmological constant does not imply a rapid expansion. In
fact the only consistent solution for $\xi\ne6$
is that the scale factor is zero, if other components, such as matter and
radiation, are assumed to be absent. 
We propose a generalized cosmological constant which does lead to 
consistent cosmology. In this case we find, remarkably, that a good fit
to high redshift supernova data is obtained purely in terms of the
generalized cosmological constant without the need for any other components
such as non-relativistic dark matter.
For the case of non-relativistic matter dominated Universe also
we do not obtain a consistent cosmic evolution if $\xi\ne 6$. 
Here again we propose a generalized mass term in the matter field action
so as to obtain a consistent cosmic evolution. In this case also 
we find that 
a good fit to the supernova data can be obtained in terms of a single 
component, namely 
non-relativistic matter, without requiring cosmological constant. 
The fit may be further improved by adding an additional component 
such as a cosmological constant. However since $\chi^2$ per degree of
freedom is less than unity purely with non-relativistic matter, there is
no motivation for such an additional component.
This is particularly interesting since in this model we may consistently
set the cosmological constant, including the generalized cosmological
constant defined in the present paper, identically equal to zero and
yet obtain consistent cosmology.
We have also shown that the model admits the 
standard Schwarzschild solution. Hence it respects the standard
tests of the Einstein's gravity on the scale of the solar system.


\begin{spacing}{1}
\begin{small}

\end{small}
\end{spacing}

\begin{thebibliography}{unsrt}


\bibitem{Anderson} 
  J.~L.~Anderson and D.~Finkelstein,
  Am.\ J.\ Phys.\  {\bf 39}, 901 (1971).

\bibitem{Einstein} A. Einstein, in {\it The Principle of Relativity}, edited by
A. Sommerfeld (Dover, New York, 1952).

\bibitem{Weinberg}
  S.~Weinberg,
  Rev.\ Mod.\ Phys.\  {\bf 61}, 1 (1989).

\bibitem{Ng1} 
  J.~J.~van der Bij, H.~van Dam and Y.~J.~Ng,
  Physica {\bf 116A}, 307 (1982).

\bibitem{Zee} 
  A.~Zee, in {\it High Energy Physics: Proceedings of the
  20th Annual Orbis Scientiae, Coral Gables, (1983)}, edited by B. Kursunoglu, S. C. Mintz, and A. Perlmutter
  (Plenum, New York, 1985). 

\bibitem{Buchmuller} 
  W.~Buchmuller and N.~Dragon,
  Phys.\ Lett.\  B {\bf 207}, 292 (1988).

\bibitem{Henneaux89}
  M.~Henneaux and C.~Teitelboim,
  Phys.\ Lett.\  B {\bf 222}, 195 (1989).

\bibitem{Unruh1989} 
  W.~G.~Unruh,
  Phys.\ Rev.\  D {\bf 40}, 1048 (1989).

\bibitem{Ng2} 
  Y.~J.~Ng and H.~van Dam,
  J.\ Math.\ Phys.\  {\bf 32}, 1337 (1991).

\bibitem{Finkelstein}
  D.~R.~Finkelstein, A.~A.~Galiautdinov and J.~E.~Baugh,
  J.\ Math.\ Phys.\  {\bf 42}, 340 (2001)
  [arXiv:gr-qc/0009099].

\bibitem{Alvarez05} 
  E.~Alvarez, 
  JHEP {\bf 0503}, 002 (2005)
  [arXiv:hep-th/0501146].

\bibitem{Alvarez06} 
  E.~Alvarez, D.~Blas, J.~Garriga, E.~Verdaguer
  Nucl.\ Phys.\  B {\bf 756}, 148 (2006)
  [arXiv:hep-th/0606019].

\bibitem{Abassi} 
  A.~H.~Abbassi and A.~M.~Abbassi,
  Class.\ Quant.\ Grav.\  {\bf 25}, 175018 (2008)
  [arXiv:0706.0451 [gr-qc]].

\bibitem{Ellis} 
  G.~F.~R.~Ellis, H.~van Elst, J.~Murugan and J.~-P.~Uzan,
  Class.\ Quant.\ Grav.\  {\bf 28}, 225007 (2011)
  [arXiv:1008.1196 [gr-qc]].


\bibitem{Alvarez07} 
  E.~Alvarez and A.~F.~Faedo,
  Phys.\ Rev.\  D {\bf 76}, 064013 (2007)
  [arXiv:hep-th/0702184].

\bibitem{JMS}
  P.~Jain, S.~Mitra and N.~K.~Singh,
  JCAP {\bf 0803}, 011 (2008)
  [arXiv:0801.2041 (astro-ph)].

\bibitem{Alvarez09} 
  E.~Alvarez, A.~F.~Faedo and J.~J.~Lopez-Villarejo,
  JCAP {\bf 0907}, 002 (2009)
  [arXiv:0904.3298 (hep-th)].

\bibitem{Alvarez10} 
  E.~Alvarez and R.~Vidal,
  Phys.\ Rev.\  D {\bf 81}, 084057 (2010)
  [arXiv:1001.4458 (hep-th)].

\bibitem{Shaposhnikov}
  D.~Blas, M.~Shaposhnikov and D.~Zenhausern,
  Phys.\ Rev.\ D {\bf 84}, 044001 (2011)
  [arXiv:1104.1392 [hep-th]].

\bibitem{Anber10} 
  M.~M.~Anber, U.~Aydemir and J.~F.~Donoghue,
  Phys.\ Rev.\  D {\bf 81}, 084059 (2010)
  [arXiv:0911.4123 (gr-qc)].
 
\bibitem{Amanullah}
  R.~Amanullah {\it et al.},
  Astrophys.\ J.\  {\bf 716}, 712 (2010)
  [arXiv:1004.1711 [astro-ph.CO]].


\end{thebibliography}
\end{document}